\documentclass[a4paper]{article}

\usepackage[margin=1in]{geometry}

\usepackage[T1]{fontenc}
\newcommand{\changefont}[3]{
\fontfamily{#1} \fontseries{#2} \fontshape{#3} \selectfont}

\changefont{ptm}{m}{n}

\usepackage{setspace} 

\usepackage{graphicx}

\usepackage{amsfonts,amsmath}
\usepackage{amssymb}
\usepackage{graphics}
\usepackage{mathrsfs}
 
\usepackage{color}

\newtheorem{theorem}{Theorem}[section]

\newtheorem{corollary}{Corollary}[section]

\newtheorem{definition}{Definition}[section]

\long\def\symbolfootnote[#1]#2{\begingroup%
\def\thefootnote{\fnsymbol{footnote}}\footnote[#1]{#2}\endgroup} 

\begin{document}

\begin{center}
\Large \textbf{Generation of Synchronous Unpredictable Oscillations by Coupled Hopfield Neural Networks}
\end{center}

\begin{center}
\normalsize \textbf{Mehmet Onur Fen$^{a,}\symbolfootnote[1]{Corresponding Author Tel.: +90 312 585 0217, E-mail: monur.fen@gmail.com}$, Fatma Tokmak Fen$^b$} \\
\vspace{0.2cm}
\textit{\textbf{\footnotesize$^a$Department of Mathematics, TED University, 06420 Ankara, Turkey}} \\
\vspace{0.1cm}
\textit{\textbf{\footnotesize$^b$Department of Mathematics, Gazi University, 06560 Ankara, Turkey}} \\
\vspace{0.1cm}
\end{center}

\vspace{0.3cm}

\begin{center}
\textbf{Abstract}
\end{center}

\noindent\ignorespaces
A new criterion based on generalized synchronization is provided for the extension of unpredictable oscillations among coupled Hopfield neural networks (HNNs). It is shown that if a drive HNN possesses an unpredictable oscillation, then a response HNN also possesses such an oscillation provided that they are synchronized in the generalized sense. Extension of unpredictability in coupled 4D HNNs are exemplified with simulations. The auxiliary system approach and conditional Lyapunov exponents are utilized to demonstrate the presence of synchronization.

\vspace{0.2cm}
 
\noindent\ignorespaces \textbf{Keywords:}  Hopfield neural networks; Generalized synchronization; Unpredictable oscillations; Auxiliary system approach; Conditional Lyapunov exponents

\vspace{0.2cm}

\noindent\ignorespaces \textbf{MSC Classification:} 34C28; 34C15; 92B25
 
\vspace{0.6cm}

\section{Introduction} \label{sec1}

In the literature, researches on chaotic dynamics starts with the studies of Poincar\'{e} on the $n$-body problem in which the sensitive dependence on initial conditions was observed \cite{Poincare57,Mawhin05}. Chaos has come into prominence with the demonstration of the phenomenon by Lorenz \cite{Lorenz63}. Nowadays, sensitive dependence on initial conditions, or in short sensitivity, is taken into account as the main ingredient \cite{Wiggins88}. The first mathematical definition of chaos was proposed by Li and Yorke \cite{Li75} based on the proximality and frequent separation of orbits. Another definition was provided by Devaney \cite{Devaney87}, and accordingly, sensitivity, density of periodic orbits, and the presence of a dense orbit are considered as the components of chaos. A collection of orbits is required in both of these mathematical descriptions.

A new type of chaos, called Poincar\'{e} chaos, was introduced in 2016 by Akhmet and Fen \cite{AkhmetUnpredictable}. It was shown in \cite{AkhmetUnpredictable} that a special type of Poisson stable trajectory, named unpredictable, leads to the presence of chaos in the quasi-minimal set. A crucial feature of Poincar\'{e} chaos is its description through a single motion, an unpredictable one, without the requirement of a collection of motions. Additionally, the sensitivity feature of motions is present when Poincar\'{e} chaos exists in a dynamics, i.e., initially nearby orbits diverge eventually. The papers \cite{AkhmetExistence}-\cite{AkhmetUnpredSolnDE} are concerned with the existence, uniqueness and stability of unpredictable oscillations in systems of differential equations. The reader is referred to the papers \cite{Akhmet20,Fen21} for unpredictable oscillations in shunting inhibitory cellular neural networks.

A phenomenon which can occur in coupled chaotic systems is synchronization \cite{Fujisaka83}-\cite{Miranda04}. The synchronization of identical chaotic systems was proposed in \cite{Pecora90} and the concept is generalized for non-identical ones by Rulkov et al. \cite{Rulkov95}. A functional relation exists between the states of the drive and response in the case that generalized synchronization (GS) occurs \cite{Rulkov95,Kocarev96}. Accordingly, it is possible to predict the dynamics of the response by means of the dynamics of the drive \cite{Miranda04,Abarbanel96}. Recently, it is rigorously proved by Tokmak Fen et al. \cite{TokmakFen22} that unidirectional coupling of two autonomous systems of differential equations leads to the extension of unpredictable oscillations provided that the drive admits such an oscillation and GS is present. 

Researches of chaos, oscillations, and synchronization are indispensable for neural networks since they take place in various neuronal activities \cite{Babloyantz96}-\cite{Erra17}. Motivated by the importance of these concepts, in the present study we handle the problem of generating synchronous unpredictable oscillations in Hopfield neural networks (HNNs) \cite{Hopfield84}, which have applications in fields such as image restoration \cite{Paik92}, object recognition \cite{Nasrabadi91}, multiuser detection \cite{Kechriotis96}, and solutions of ill-posed problems \cite{Tavares21}. For that purpose, we take into account unidirectionally coupled HNNs in which the drive is known to admit an unpredictable oscillation, and analyze the dynamics of the response. The drive admits a special type of input, generated via the logistic map. Based on the theoretical achievements provided in papers \cite{AkhmetUnpredSolnDE} and \cite{TokmakFen22}, we demonstrate that the response also possesses an unpredictable oscillation, which is in synchrony with the one of the drive, in the case that GS occurs \cite{Rulkov95}. The continuity of the function which relates the outputs of the networks is required, and that criterion is guaranteed by the auxiliary system approach \cite{Abarbanel96}. We utilize this approach to detect GS in the coupled HNNs. Other techniques which can also be used for the same purpose are mutual false nearest neighbors, conditional Lyapunov exponents, and Lyapunov functions \cite{Rulkov95,Miranda04,Kocarev96,He92}. Our technique provides a way to design Hopfield type neural systems possessing unpredictable oscillations.

The existence of unpredictable oscillations in HNNs was demonstrated in \cite{AkhmetUnpredSolnDE} benefiting from continuous unpredictable external inputs, and the extension of such oscillations was also discussed. The approach of the present study is different compared to \cite{AkhmetUnpredSolnDE} such that the extension of unpredictability is achieved whenever GS occurs in the dynamics. For that reason, according to our novel approach it is possible to generate unpredictable oscillations in a larger class of HNNs, which can display various dynamical phenomena when the driving is not established. In the absence of the driving, the response HNN can admit not only point attractors but also limit cycles or chaotic attractors as well. Moreover, the smallness of connection weights and Lipschitz constants are not required for the response network. The example given in Section \ref{sec4} reveals that the response can possess even hyperchaos in the absence of the driving. However, due to the smallness of the Lipschitz constants of the activation functions and connection weights, such phenomena cannot take place in the class of HNNs handled in \cite{AkhmetUnpredSolnDE}, and GS was not considered in that study.

\section{Preliminaries} \label{sec2}

Throughout the paper $\mathbb R$ and $\mathbb R^m$ respectively stand for the sets of real numbers and $m\times 1$ real column vectors, $(\cdot)^T$ symbolizes transpose, and $\left\|\cdot\right\|$ denotes the Euclidean vector norm.

Dynamics of a HNN can be described by the nonlinear system
\begin{eqnarray} \label{HNN1}
\displaystyle \frac{dx_i}{dt} = - c_i x_i + \displaystyle \sum_{j=1}^{m} w_{ij} f_j(x_j) + I_i, ~~ i=1,2,\ldots,m,
\end{eqnarray}
where $m$ is the number of neurons, $x_i$ is the total input to $i$th neuron, the bounded monotonic differentiable function $f_j:\mathbb R \to \mathbb R$ is the activation function acting on $j$th neuron, $c_i>0$  are real constants, $I_i$ is the external input of the $i$th neuron, and $w_{ij}$ is the synaptic connection weight between the $i$th and $j$th neurons \cite{Hopfield84,Peng05,Mathias,Li13,Bao17}. 

In this paper, we investigate coupled HNNs with a skew product structure in which the drive is of the form (\ref{HNN1}) and the dynamics of the response HNN is governed by the nonlinear ordinary differential equations 
\begin{eqnarray} \label{HNN2}
\displaystyle \frac{dy_i}{dt} = - \widetilde{c}_i y_i + \displaystyle \sum_{j=1}^{n} \widetilde{w}_{ij} g_j(y_j) + h_i(x(t)), ~ i=1,2,\ldots,n,
\end{eqnarray}
where $n$ is the number of neurons, $x(t)=(x_1(t),x_2(t),\ldots, x_m(t))^T$ is an output of the drive (\ref{HNN1}), for each $i$ the functions $h_i: \mathbb R \to \mathbb R$ are continuous, and the constants $\widetilde{c}_i>0$, the connection weights $\widetilde{w}_{ij}$ as well as the activation functions $g_j:\mathbb R \to \mathbb R$ are respectively the counterparts of $c_i$, $w_{ij}$, $f_j$ in (\ref{HNN1}).

We mainly assume that the HNNs (\ref{HNN1}) and (\ref{HNN2}) admit compact invariant sets $\Lambda_x \subset \mathbb R^m$ and $\Lambda_y \subset \mathbb R^n$, respectively. Accordingly, an output of the coupled network (\ref{HNN1})-(\ref{HNN2}) which starts in   $\Lambda_x \times \Lambda_y$ remains in the same set for $t \geq 0$.

We consider a class of GS in which the state $y(t)=(y_1(t),y_2(t),\ldots,y_n(t))^T$ of the response (\ref{HNN2}) is a continuous function of the state $x(t)=(x_1(t),x_2(t),\ldots,x_m(t))^T$ of the drive (\ref{HNN1}). We say that GS occurs in the dynamics of the coupled network (\ref{HNN1})-(\ref{HNN2}) if there is a continuous transformation $\psi$ such that for each $\alpha_0 \in \Lambda_x$, $\beta_0 \in \Lambda_y$ the relation 
\begin{eqnarray} \label{synchrelation}
\displaystyle \lim_{t\to\infty} \left\|y(t)-\psi(x(t))\right\|=0 
\end{eqnarray} 
holds, where $x(t)$ and $y(t)$ are respectively the outputs of (\ref{HNN1}) and (\ref{HNN2}) satisfying $x(0)=\alpha_0$ and $y(0)=\beta_0$. Further information related to the continuity of the transformation $\psi$ can be found in the studies \cite{Miranda04,Abarbanel96}.

The description of an unpredictable sequence is given in the next definition.
\begin{definition} (\cite{AkhmetUnpredSolnDE}) \label{unpseqquasi}
A bounded sequence $\left\{\lambda^*_k\right\}_{k \in \mathbb Z}$ is called unpredictable if there exist a positive number $\delta_{0}$  and sequences $\left\{\zeta_n\right\}_{n\in\mathbb N},$ $\left\{\eta_n\right\}_{n\in\mathbb N}$ of positive integers both of which diverge to infinity such that $\big\|\lambda^*_{k+\zeta_n} - \lambda^*_k \big\|\to 0$ as $n \to \infty$ for each $k$ in bounded intervals of integers and $ \big\| \lambda^*_{\zeta_n + \eta_n} - \lambda^*_{\eta_n}\big\| \geq \delta_{0}$ for each $n\in\mathbb N.$
\end{definition}

The definition of an unpredictable function is as follows.

\begin{definition} (\cite{AkhmetUnpredSolnDE}) \label{unpoutput}
A uniformly continuous and bounded function $\phi(t)=(\phi_1(t), \phi_2(t), \ldots, \phi_m(t))^T$, where $\phi_i(t)$ are real valued functions for each $i=1,2,\ldots,m$, is called unpredictable if there exist positive numbers $\epsilon_0$, $r$ and sequences $\left\{\mu_k\right\}_{k\in\mathbb N}$ and $\left\{\nu_k\right\}_{k\in\mathbb N}$ both of which diverge to infinity such that $\left\|\phi(t+\mu_k)-\phi(t)\right\| \to 0$ as $k \to \infty$ uniformly on compact subsets of $\mathbb R$ and $\left\|\phi(t+\mu_k)-\phi(t)\right\| \geq \epsilon_0$ for each $t \in [\nu_k-r,\nu_k+r]$ and $k \in \mathbb N.$
\end{definition}

\section{Theoretical Foundations} \label{sec3}
In this section we provide an application of the theoretical results given in papers \cite{AkhmetUnpredSolnDE} and \cite{TokmakFen22} to unidirectionally coupled HNNs.

First of all, let us consider the logistic map 
\begin{eqnarray} \label{logisticmap}
\lambda_{k+1}=\gamma \lambda_k (1-\lambda_k),
\end{eqnarray} 
where $k \in \mathbb Z$ and $\gamma$ is a real parameter. For the values of the parameter $\gamma$ between $3+(2/3)^{1/2}$ and $4$ the map (\ref{logisticmap}) admits an unpredictable orbit \cite{AkhmetPoincare}, and the unit interval $[0,1]$ is invariant under its iterations \cite{Hale91}.

For a fixed value of the parameter $\gamma$ from the interval $[3+(2/3)^{1/2},4]$, let us denote by $\left\{\lambda^*_k\right\}_{k \in \mathbb Z}$ an unpredictable orbit of (\ref{logisticmap}), which belongs to the interval $[0,1]$, and define the function $\Omega:\mathbb R \to \mathbb R$ through the equation 
\begin{eqnarray}\label{omega1}
\Omega(t)=\lambda^*_k  
\end{eqnarray} 
for $t \in [k,k+1)$ and $k \in \mathbb Z.$
The function $\Omega(t)$ is the solution of the impulsive differential equation
\begin{eqnarray} \label{impulsive}
\begin{array}{l} \displaystyle \frac{d\Omega}{dt}=0,  \ t \neq k,  \\
 \Delta \Omega |_{t=k}=\lambda^*_{k}-\lambda^*_{k-1} 
\end{array} \end{eqnarray}
with $\Omega(0)=\lambda^*_{-1}$, where $\Delta \Omega |_{t=k}=\Omega(k+)-\Omega(k)$ and $\Omega(k+)=\displaystyle \lim_{t \to k^+} \Omega(t)$. Additionally, $\Omega(t)$ admits the same impulse moments with the solution of the discontinuous autonomous dynamical system 
\begin{eqnarray} \label{impmoments}
\begin{array}{l}\displaystyle \frac{dz}{dt}=-1,  \\ 
\Delta z |_{z=0}=1 
\end{array} \end{eqnarray}
satisfying   $z(0)=0$. 

Using the arguments presented in Section 4 of paper \cite{AkhmetUnpredSolnDE}, one can confirm that for any fixed positive number $\alpha$ the function 
\begin{eqnarray}\label{functionH} 
H(t)=\displaystyle \int^t_{-\infty} e^{-\alpha (t-s)} \Omega(s) ds
\end{eqnarray} 
is unpredictable, and  it is the unique asymptotically stable bounded solution of the differential equation
\begin{eqnarray}\label{bddsoln} 
\displaystyle\frac{dv}{dt} = -\alpha v + \Omega(t).
\end{eqnarray}

Next, we suppose that the external inputs $I_i$, $i=1,2,\ldots,m$, in (\ref{HNN1}) are defined by means of the function $H(t)$. More precisely, we take into account the drive HNN of the form 
\begin{eqnarray} \label{HNNdrive}
\displaystyle \frac{dx_i}{dt} = - c_i x_i + \displaystyle \sum_{j=1}^{m} w_{ij} f_j(x_j) + d_iH(t), ~ i=1,2,\ldots,m,
\end{eqnarray}
where $H(t)$ is the unpredictable function defined by (\ref{functionH}) and at least one of the real constants $d_i$, $i=1,2,\ldots,m$, is nonzero. In network (\ref{HNNdrive}), $x_i$, $c_i$, $w_{ij}$, and $f_j$ have the same meanings as in (\ref{HNN1}). According to Theorem 5.2 \cite{AkhmetPoincare}, the function $\Theta: \mathbb R \to \mathbb R^n$ defined by 
\begin{eqnarray*} \label{unpinput}
\Theta(t)=(d_1 H(t), d_2 H(t), \ldots, d_n H(t))^T
\end{eqnarray*}
is also unpredictable. It is worth noting that the network (\ref{logisticmap})-(\ref{impulsive})-(\ref{impmoments})-(\ref{bddsoln})-(\ref{HNNdrive}) is a hybrid one since its dynamics is governed by both differential and discrete equations, and it is autonomous.

The following assumptions on the networks (\ref{HNNdrive}) and (\ref{HNN2}) are required. 
\begin{itemize}
\item[\textbf{(A1)}] There exist positive numbers $M_j$, $j=1,2,\ldots,m$, such that $\left| f_j(u) \right| \leq M_j$ for each $u\in\mathbb R$;
\item[\textbf{(A2)}] There exist positive numbers $l_j$, $j=1,2,\ldots,m$, such that $\left\|f_j(u) - f_j(\overline{u}) \right\| \leq l_j \left\|u-\overline{u}\right\|$ for each $u, \overline{u} \in \mathbb R$; 
\item[\textbf{(A3)}] There exist positive numbers $\widetilde{l}_j$, $j=1,2,\ldots,n$, such that $\left\|g_j(u) - g_j(\overline{u}) \right\| \leq \widetilde{l}_j \left\|u-\overline{u}\right\|$ for each $u, \overline{u} \in \mathbb R$; 
\item[\textbf{(A4)}] There exists a positive number $l$ such that $\left\|h(x) - h(\overline{x}) \right\| \geq l \left\|x-\overline{x}\right\|$ for each $x,\overline{x} \in\Lambda_x$, where $h(x)=(h_1(x),h_2(x),\ldots,h_n(x))^T$.
\end{itemize}

One can confirm using Theorem 2.1 \cite{AkhmetUnpredSolnDE} that if the assumptions $(A1)$ and $(A2)$ hold, then the HNN (\ref{HNNdrive}) admits a unique uniformly exponentially stable unpredictable output, provided that $\left|w_{ij}\right|$ and $l_j$, $i,j=1,2,\ldots,m$, are sufficiently small. 
 
Based on Theorem 3.1 given in paper \cite{TokmakFen22}, we have the following result for the response HNN (\ref{HNN2}).

\begin{theorem} \label{thm2}
Suppose that the assumptions $(A1)-(A4)$ are fulfilled.  If GS takes place in the dynamics of the coupled network (\ref{HNNdrive})-(\ref{HNN2}), then the response HNN (\ref{HNN2}) possesses an unpredictable output, provided that $\left|w_{ij}\right|$ and $l_j$, $i,j=1,2,\ldots,m$, are sufficiently small.
\end{theorem}	
 
The following corollary of Theorem \ref{thm2}  can be verified using Remark 3.1 \cite{TokmakFen22}.
\begin{corollary} \label{corollary1}
The coupled network (\ref{HNNdrive})-(\ref{HNN2}) admits an unpredictable output under the conditions of Theorem \ref{thm2}. 
\end{corollary}
 
The next section is devoted to the demonstration of synchronous unpredictable oscillations in coupled $4$-neuron HNNs.

\section{Unpredictable Behavior of Coupled 4D HNNs} \label{sec4}
 
Let us use the parameter $\gamma=3.93$ in the logistic map (\ref{logisticmap}) and define $\Omega(t)$ by (\ref{omega1}), in which $\left\{\lambda_k^*\right\}_{k \in \mathbb Z}$ is an unpredictable orbit of (\ref{logisticmap}). We consider the $4$D HNN 
\begin{eqnarray} \label{example1}
\begin{array}{l}
 \displaystyle \frac{dx_1}{dt} = -0.3 x_1 + 0.02 \tanh(x_1) - 0.06 \tanh(x_2) -0.05 \tanh(x_3)+ 0.03 \tanh(x_4) +5.4 H(t) \\
 \displaystyle \frac{dx_2}{dt} = -0.6 x_2 + 0.01 \tanh(x_1) + 0.04 \tanh(x_2) + 0.06 \tanh(x_3) +0.02 \tanh(x_4) -2.6 H(t)\\
 \displaystyle \frac{dx_3}{dt} = -0.8 x_3 - 0.03 \tanh(x_1) + 0.08 \tanh(x_2) + 0.04 \tanh(x_3) +0.04 \tanh(x_4) -1.2 H(t)\\
 \displaystyle \frac{dx_4}{dt} = -0.5 x_4 + 0.07 \tanh(x_1) - 0.01 \tanh(x_2) + 0.02 \tanh(x_3) +0.05 \tanh(x_4) + 3.8H(t),
\end{array}\end{eqnarray}
where $H(t)$ is the continuous function defined by (\ref{functionH}) with  $\alpha=3$.

HNN (\ref{example1}) is in the form of (\ref{HNNdrive}) with $m=4$, $c_1=0.3$, $c_2=0.6$, $c_3=0.8$, $c_4=0.5$, $d_1=5.4$, $d_2=-2.6$, $d_3=-1.2$, $d_4=3.8$, $f_j(x_j)=\tanh(x_j)$ for each $j=1,2,3,4$, and
\[
\begin{pmatrix}  
  w_{11} & w_{12} &  w_{13} &  w_{14} \\ 
  w_{21} & w_{22} &  w_{23} &  w_{24} \\ 
  w_{31} & w_{32} &  w_{33} &  w_{34} \\ 
  w_{41} & w_{42} &  w_{43} &  w_{44} 
\end{pmatrix}
=
\begin{pmatrix}  
  0.02 & - 0.06 &  -0.05 & 0.03 \\ 
  0.01 & 0.04 &  0.06 &  0.02 \\ 
  - 0.03 & 0.08 &  0.04 &  0.04  \\ 
 0.07 & - 0.01 &   0.02  &  0.05
\end{pmatrix}.
\]
One can verify that the assumptions $(A1)$ and $(A2)$ are valid for (\ref{example1}). The connection weights $w_{ij}$, $i,j=1,2,3,4$, in (\ref{example1}) are sufficiently small in absolute value such that the network possesses a unique uniformly exponentially stable unpredictable output in accordance with Theorem 2.1 \cite{AkhmetUnpredSolnDE}.

\begin{figure}[ht!] 
\centering
\includegraphics[width=16.0cm]{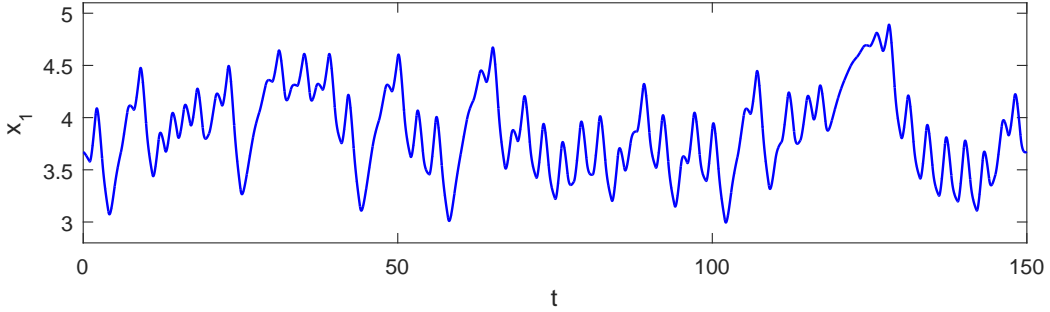}
\caption{The irregular behavior of the $x_1$-coordinate of the hybrid network (\ref{logisticmap})-(\ref{impulsive})-(\ref{impmoments})-(\ref{bddsoln})-(\ref{example1}). The initial data  $\lambda_0=0.48$, $v(0)=0.18$, $x_1(0)=3.67$, $x_2(0)=-0.79$, $x_3(0)=-0.34$, and $x_4(0)=1.59$ are used in the simulation. The irregular behavior approves the presence of an unpredictable oscillation in the dynamics of (\ref{example1}).}
\label{fig1}
\end{figure}

To visualize the unpredictable behavior of (\ref{example1}), in Figure \ref{fig1} we show the time-series of the $x_1$-coordinate of the autonomous hybrid network (\ref{logisticmap})-(\ref{impulsive})-(\ref{impmoments})-(\ref{bddsoln})-(\ref{example1}) corresponding to the initial data $\lambda_0=0.48$, $v(0)=0.18$, $x_1(0)=3.67$, $x_2(0)=-0.79$, $x_3(0)=-0.34$, and $x_4(0)=1.59$.          

Next, we take into account the $4$D HNN \cite{Li05}
\begin{eqnarray} \label{example2}
\begin{array}{l}
 \displaystyle \frac{dy_1}{dt} = - y_1 +  \tanh(y_1) + 0.5 \tanh(y_2) - 3 \tanh(y_3) -\tanh(y_4) \\
 \displaystyle \frac{dy_2}{dt} = - y_2 +  2.3 \tanh(y_2) + 3 \tanh(y_3) \\
 \displaystyle \frac{dy_3}{dt} = - y_3 +3 \tanh(y_1) - 3\tanh(y_2) + \tanh(y_3) \\
 \displaystyle \frac{dy_4}{dt} = -100 y_4 + 100 \tanh(y_1)  +170 \tanh(y_4).
\end{array}\end{eqnarray} 
 It was demonstrated by Li et al. \cite{Li05} that the network (\ref{example2}) is hyperchaotic such that its Lyapunov exponents are $0.237$, $0.024$, $0$, $-74.08$ and the Lyapunov dimension is $3.002$.
 
In order to extend the unpredictable behavior, we take HNN (\ref{example1}) as the drive and establish unidirectional coupling between (\ref{example1}) and (\ref{example2}) by setting up the response HNN
\begin{eqnarray} \label{example3}
\begin{array}{l}
 \displaystyle \frac{dy_1}{dt} = - y_1 +  \tanh(y_1) + 0.5 \tanh(y_2) - 3 \tanh(y_3) -\tanh(y_4) + 8.5x_1(t)\\
 \displaystyle \frac{dy_2}{dt} = - y_2 +  2.3 \tanh(y_2) + 3 \tanh(y_3) + 6.8x_2(t)\\
 \displaystyle \frac{dy_3}{dt} = - y_3 +3 \tanh(y_1) - 3\tanh(y_2) + \tanh(y_3) + 9.4x_3(t)\\
 \displaystyle \frac{dy_4}{dt} = -100 y_4 + 100 \tanh(y_1)  +170 \tanh(y_4)+ 7.1x_4(t),
\end{array}\end{eqnarray}  
where $x(t)=(x_1(t),x_2(t),x_3(t),x_4(t))^T$ is an output of (\ref{example1}).
 
HNN (\ref{example3}) is in the form of (\ref{HNN2}) with $n=4$, $\widetilde c_1= \widetilde c_2= \widetilde c_3=1$, $\widetilde c_4=100$, $g_j(y_j)=\tanh(y_j)$ for each $j=1,2,3,4$, $h(x_1,x_2,x_3,x_4)=(8.5x_1, 6.8x_2, 9.4x_3, 7.1x_4)^T$, and
\[
\begin{pmatrix}  
 \widetilde  w_{11} & \widetilde  w_{12} & \widetilde  w_{13} & \widetilde  w_{14} \\ 
 \widetilde  w_{21} & \widetilde  w_{22} & \widetilde  w_{23} & \widetilde  w_{24} \\ 
 \widetilde  w_{31} & \widetilde w_{32} & \widetilde  w_{33} & \widetilde  w_{34} \\ 
 \widetilde  w_{41} & \widetilde  w_{42} & \widetilde  w_{43} & \widetilde  w_{44} 
\end{pmatrix}
=
\begin{pmatrix}  
 1 & 0.5 &  -3 & -1 \\ 
  0  & 2.3 &  3 &  0 \\ 
  3 & -3 & 1 &  0   \\ 
 100 & 0 &   0  &  170
\end{pmatrix}.
\] 
It can be confirmed that the assumptions $(A3)$ and $(A4)$ are satisfied for (\ref{example3}).
 
Now, we will demonstrate that GS takes place in the dynamics of the coupled network (\ref{example1})-(\ref{example3}). For that purpose, let us make use of the auxiliary system approach \cite{Abarbanel96}. Consider the HNN
 \begin{eqnarray} \label{example4}
\begin{array}{l}
 \displaystyle \frac{dz_1}{dt} = - z_1 +  \tanh(z_1) + 0.5 \tanh(z_2) - 3 \tanh(z_3) -\tanh(z_4) + 8.5x_1(t)\\
 \displaystyle \frac{dz_2}{dt} = - z_2 +  2.3 \tanh(z_2) + 3 \tanh(z_3) + 6.8x_2(t)\\
 \displaystyle \frac{dz_3}{dt} = - z_3 +3 \tanh(z_1) - 3\tanh(z_2) + \tanh(y_3) + 9.4x_3(t)\\
 \displaystyle \frac{dz_4}{dt} = -100 z_4 + 100 \tanh(z_1)  +170 \tanh(z_4)+ 7.1x_4(t),
\end{array}\end{eqnarray}  
which is an identical copy of the response (\ref{example3}). 

\begin{figure}[ht!] 
\centering
\includegraphics[width=10.0cm]{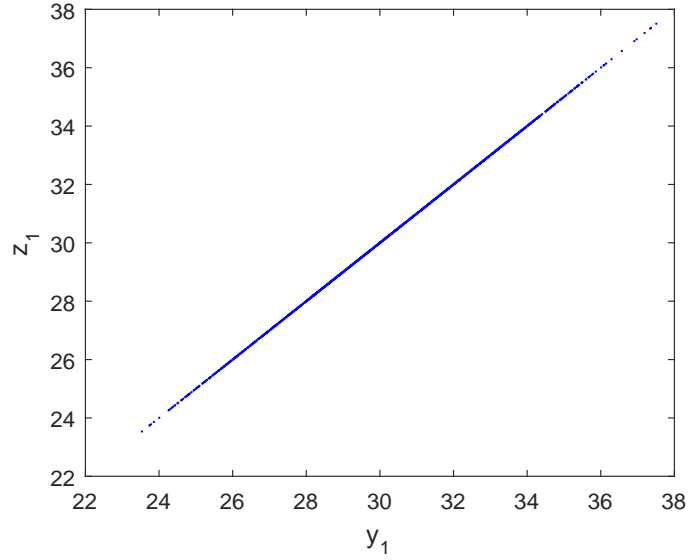}
\caption{The result of the auxiliary system approach utilized for the coupled HNNs (\ref{example1})-(\ref{example3}). Since the plot takes place on the line $z_1=y_1$ GS is present in the dynamics.}
\label{fig2}
\end{figure}

In the case that we use the output $x(t)=(x_1(t), x_2(t), x_3(t),x_4(t))^T$ of (\ref{example1}) whose first coordinate is represented in Figure \ref{fig1}, we obtain the projection of the stroboscopic plot of the network (\ref{example1})-(\ref{example3})-(\ref{example4}) on the $y_1-z_1$ shown in Figure \ref{fig2}. In the simulation we benefit from the initial data $y_1(0)= 28.87$, $y_2(0)= -5.09 $, $y_3(0)=3.65$, $y_4(0)=2.79$, $z_1(0)= 24.28$, $z_2(0)= -6.24$, $z_3(0)=4.59$, $z_4(0)=3.26$. The first $200$ iterations are omitted to eliminate the transients. Because the plot takes place on the line $z_1=y_1$, GS is present in the dynamics of the coupled network (\ref{example1})-(\ref{example3}). In other words, the relation (\ref{synchrelation}) is fulfilled for some continuous transformation $\psi$ \cite{Miranda04,Abarbanel96}, in which $x(t)=(x_1(t), x_2(t), x_3(t),x_4(t))^T$ and $y(t)=(y_1(t), y_2(t), y_3(t),y_4(t))^T$ refer to the outputs of (\ref{example1}) and (\ref{example3}), respectively.
 
Now, we will affirm the occurrence of GS one more time by means of the conditional Lyapunov exponents. Let us constitute the following variational system for (\ref{example3}):
\begin{eqnarray} \label{variational} \begin{array}{l}
\displaystyle \frac{d \xi_1}{dt} = \left[ -1+ \textrm{sech}^2(y_1(t))\right]  \xi_1 + 0.5 ~\textrm{sech}^2(y_2(t)) \xi_2 - 3~ \textrm{sech}^2(y_3(t)) \xi_3 - \textrm{sech}^2(y_4(t)) \xi_4 \\
\displaystyle \frac{d \xi_2}{dt} =  \left[-1+ 2.3 ~\textrm{sech}^2(y_2(t)) \right] \xi_2 + 3 ~ \textrm{sech}^2(y_3(t)) \xi_3 \\
\displaystyle \frac{d \xi_3}{dt} = 3 ~\textrm{sech}^2(y_1(t)) \xi_1 -3 ~\textrm{sech}^2(y_2(t)) \xi_2 +  \left[-1+ \textrm{sech}^2(y_3(t)) \right] \xi_3 \\
\displaystyle \frac{d \xi_4}{dt} = 100 ~\textrm{sech}^2(y_1(t)) \xi_1  +  \left[-100+ 170 ~ \textrm{sech}^2(y_4(t)) \right] \xi_4. 
\end{array} \end{eqnarray} 
When the output $y(t)=(y_1(t),y_2(t),y_3(t),y_4(t))^T$ of (\ref{example3}) corresponding
to the initial data $\lambda_0=0.48$, $v(0)=0.18$, $x_1(0)=3.67$, $x_2(0)=-0.79$, $x_3(0)=-0.34$, $x_4(0)=1.59$, $y_1(0)= 28.87$, $y_2(0)=-5.09$, $y_3(0)= 3.65$, $y_4(0)= 2.79$ is used, the largest Lyapunov exponent of (\ref{variational}) is evaluated as $-0.9968$. All conditional Lyapunov exponents of the response HNN (\ref{example3}) are hence negative, and this certifies that the HNNs (\ref{example1}) and (\ref{example3}) are synchronized in the generalized sense. Therefore, the response HNN (\ref{example3}) admits an unpredictable oscillation according to Theorem \ref{thm2}, and it is in synchrony with the one of the drive by equation (\ref{synchrelation}). Additionally, the coupled network (\ref{example1})-(\ref{example3}) also possesses such an oscillation by Corollary \ref{corollary1}.

\begin{figure}[ht!] 
\centering
\includegraphics[width=16.0cm]{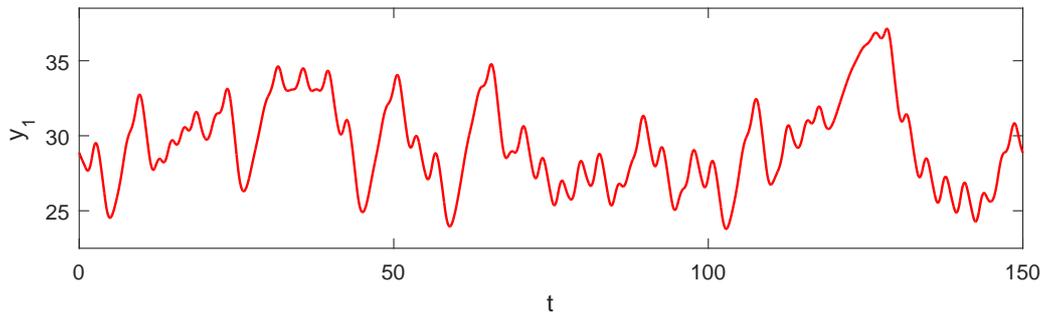}
\caption{The irregular behavior of the $y_1$-coordinate of the coupled HNNs (\ref{example1})-(\ref{example3}). The presented time-series corresponds to the initial data  $y_1(0)= 28.87$, $y_2(0)=-5.09  $, $y_3(0)= 3.65$, $y_4(0)= 2.79$. The irregular behavior observed in the figure reveals the existence of an unpredictable oscillation in the dynamics of (\ref{example3}).}
\label{fig3}
\end{figure}

Utilizing again the output $x(t)=(x_1(t),x_2(t),x_3(t),x_4(t))^T$ of (\ref{example1}) whose first coordinate is depicted in Figure \ref{fig1}, we represent the time-series of the first neuron, $y_1$, of (\ref{example3}) in Figure \ref{fig3}. The initial data  $y_1(0)= 28.87$, $y_2(0)=-5.09  $, $y_3(0)= 3.65$, $y_4(0)= 2.79$ are used. The irregular behavior shown in Figure \ref{fig3} confirms the result of Theorem \ref{thm2} such that an unpredictable oscillation of (\ref{example1}) is extended.

\section{Conclusion}

In this study, we show that if GS is present in coupled HNNs, then an unpredictable oscillation can be inherited from the drive. In other words, the drive and response networks possess synchronous unpredictable oscillations such that the oscillation of the latter is a continuous image of the one of the former. The response HNN can display different dynamical behavior such as point attractors, limit cycles, chaotic attractors, or even hyperchaos in the absence of the driving. When the driving is established, Theorem \ref{thm2} imply that the response, as well as the coupled network, necessarily admit unpredictable oscillations, which are in synchrony with the one of the drive. The auxiliary system approach \cite{Abarbanel96} and the technique of conditional Lyapunov exponents \cite{Kocarev96,He92} are convenient to verify GS in the dynamics. Our results provide an opportunity to design HNNs with arbitrary high number of neurons. In the future, results based on Lyapunov functions \cite{Yoshizawa75} can be obtained. Moreover, the proposed technique can be developed for fractional-order HNNs \cite{Batiha20,He21}.


\begin{thebibliography}{30}

\bibitem{Poincare57} H. Poincar\'{e}, Les Methodes Nouvelles de la Mecanique Celeste, Vol. I, II, III, Paris, 1899; reprint, Dover, New York, 1957.

\bibitem{Mawhin05} J. Mawhin, Henri Poincar\'{e}. A life in the service of science, Notices of the AMS \textbf{52} (2005) 1032-1040.

\bibitem{Lorenz63} E. N. Lorenz, Deterministic nonperiodic flow, Journal of the Atmospheric Sciences \textbf{20} (1963) 130--141.

\bibitem{Wiggins88} S. Wiggins, Global Bifurcations and Chaos, Springer, New York, 1988.

\bibitem{Li75} T. Y. Li, J. A. Yorke, Period three implies chaos, American Mathematical Monthly \textbf{82} (1975) 985-992.	

\bibitem{Devaney87} R. Devaney, An Introduction to Chaotic Dynamical Systems, United States of America, Addison-Wesley, 1987.

\bibitem{AkhmetUnpredictable} M. Akhmet, M. O. Fen, Unpredictable points and chaos, Communications in Nonlinear Science and Numerical Simulation \textbf{40} (2016) 1-5.

\bibitem{AkhmetExistence} M. Akhmet, M. O. Fen, Existence of unpredictable solutions and chaos, Turkish Journal of Mathematics \textbf{41} (2017) 254-266.
 
\bibitem{AkhmetPoincare} M. Akhmet, M. O. Fen, Poincar\'{e} chaos and unpredictable functions, Communications in Nonlinear Science and Numerical Simulation \textbf{48} (2017) 85-94. 

\bibitem{AkhmetUnpredSolnDE} M. Akhmet, M. O. Fen, Non-autonomous equations with unpredictable solutions, Communications in Nonlinear Science and Numerical Simulation \textbf{59} (2018) 657-670.

\bibitem{Akhmet20} M. Akhmet, R. Seilova, M. Tleubergenova, A. Zhamanshin, Shunting inhibitory cellular neural networks with strongly unpredictable oscillations, Communications in Nonlinear Science and Numerical Simulation \textbf{89} (2020) 105287.

\bibitem{Fen21} M. O. Fen, F. Tokmak Fen, Unpredictable oscillations of SICNNs with delay, Neurocomputing \textbf{464} (2021) 119-129.

\bibitem{Fujisaka83} H. Fujisaka, T. Yamada, Stability theory of synchronized motion in coupled-oscillator systems, Progress of Theoretical Physics \textbf{69} (1983) 32-47.

\bibitem{Afraimovich86} V. S. Afraimovich, N. N. Verichev, M. I. Rabinovich, Stochastic synchronization of oscillation in dissipative systems, Radiophysics and Quantum Electronics \textbf{29} (1986) 795-803.

\bibitem{Pecora90} L. M. Pecora, T. L. Carroll, Synchronization in chaotic systems, Physical Review Letters \textbf{64} (1990) 821-825.

\bibitem{Rulkov95}  N. F. Rulkov, M. M. Sushchik, L. S. Tsimring, H. D. I. Abarbanel, Generalized synchronization of chaos in directionally coupled chaotic systems, Physical Review E \textbf{51} (1995) 980-994.	

\bibitem{Miranda04} J. M. Gonz\'{a}les-Miranda, Synchronization and Control of Chaos, Imperial College Press, London, 2004.

\bibitem{Kocarev96} L. Kocarev, U. Parlitz, Generalized synchronization, predictability, and equivalence of unidirectionally coupled dynamical systems, Physical Review Letters \textbf{76} (1996) 1816-1819.		
	
\bibitem{Abarbanel96} H. D. I. Abarbanel, N. F. Rulkov, M. M. Sushchik, Generalized synchronization of chaos: the auxiliary system approach, Physical Review E \textbf{53} (1996)	4528-4535.

\bibitem{TokmakFen22} F. Tokmak Fen, M. O. Fen, M. Akhmet, A novel criterion for unpredictable motions, arXiv: 2202.02422v1 (submitted).

\bibitem{Babloyantz96} A. Babloyantz, C. Lourenco, Brain chaos and computation, International Journal of Neural Systems \textbf{7} (1996) 461-471.

\bibitem{Korn03} H. Korn, P. Faure, Is there chaos in the brain? II. Experimental evidence and related models, Comptes Rendus Biologies \textbf{326} (2003) 787-840.

\bibitem{Doelling21} K. B. Doelling, M. F. Assaneo, Neural oscillations are a start toward understanding brain activity rather the end, Plos Biology \textbf{19} e3001234.

\bibitem{Zeng14} X. Zeng, Q. Hui, W. M. Haddad, T. Hayakawa, J. M. Bailey, Synchronization of biological neural network systems with stochastic perturbations and time delays, Journal of the Franklin Institute \textbf{351} (2014) 1205-1225.

\bibitem{Gray94} C. M. Gray, Synchronous oscillations in neural systems: Mechanisms and functions, Journal of Computational Neuroscience \textbf{1} (1994) 11-38.

\bibitem{Erra17} R. G. Erra, J. L. P. Velazquez, M. Rosenblum, Neural synchronization from the perspective of non-linear dynamics, Frontiers in Computational Neuroscience (2017). Doi: 10.3389/fncom.2017.00098

\bibitem{Hopfield84} J. J. Hopfield, Neurons with graded response have collective computational properties like those of two-state neurons, Proc. Natl. Acad. Sci. \textbf{81} (1984) 3088-3092.

\bibitem{Paik92} J. K. Paik, A. K. Katsaggelos, Image restoration using a modified Hopfield network, IEEE Transactions on Image Processing \textbf{1} (1992) 49-63.

\bibitem{Nasrabadi91} N. M. Nasrabadi, W. Li, Object recognition by a Hopfield neural network, IEEE Transactions on Systems, Man, and Cybernetics \textbf{21} (1991) 1523-1535.

\bibitem{Kechriotis96} G. I. Kechriotis, E. S. Manolakos, Hopfield neural network implementation of the optimal CDMA multiuser detector, IEEE Transactions on Neural Networks \textbf{7} (1996) 131-141.

\bibitem{Tavares21} C. A. Tavares, T. M. R. Santos, N. H. T. Lemes, J. P. C. dos Santos, J. C. Ferreira, J. P. Braga, Solving ill-posed problems faster using fractional-order Hopfield neural network, Journal of Computational and Applied Mathematics \textbf{381} (2021) 112984.

\bibitem{He92} R. He, P. G. Vaidya, Analysis and synthesis of synchronous periodic and chaotic systems, Physical Review A \textbf{46} (1992) 7387-7392.

\bibitem{Peng05} J. Peng, Z.-B. Xu, H. Qiao, A critical analysis on global convergence of Hopfield-type neural networks, IEEE Transactions on Circuits and Systems-I: Regular Papers \textbf{52} (2005) 804-814.

\bibitem{Mathias} A. C. Mathias, P. C. Rech, Hopfield neural network: The hyperbolic tangent and the piecewise-linear activation functions, Neural Networks \textbf{34} (2012) 42-45.

\bibitem{Li13} J. Li, F. Liu, Z.-H. Guan, T. Li, A new chaotic Hopfield neural network and its synthesis via parameter switchings, Neurocomputing \textbf{117} (2013) 33-39.

\bibitem{Bao17} B. Bao, H. Qian, J. Wang, Q. Xu, M. Chen, H. Wu, Y. Yu, Numerical analyses and experimental validations of coexisting multiple attractors in Hopfield neural network, Nonlinear Dynamics \textbf{90} (2017) 2359-2369.

\bibitem{Hale91} J. Hale, H. Ko\c{c}ak, Dynamics and Bifurcations, New York, Springer-Verlag, 1991. 

\bibitem{Li05} Q. Li, X.-S. Yang, F. Yang, Hyperchaos in Hopfield-type neural networks, Neurocomputing \textbf{67} (2005) 275-280.

\bibitem{Yoshizawa75} T. Yoshizawa, Stability Theory and the Existence of Periodic Solutions and Almost Periodic Solutions, Springer-Verlag, New-York, Heidelberg, Berlin, 1975. 

\bibitem{Batiha20} I. M. Batiha, R. B. Albadarneh, S. Momani, I. H. Jebril, Dynamics analysis of fractional-order Hopfield neural networks, International Journal of Biomathematics \textbf{13} (2020) 2050083.

\bibitem{He21} B.-B. He, H.-C. Zhou, Asymptotic stability and synchronization of fractional order Hopfield neural networks with unbounded delay, Mathematical Methods in the Applied Sciences (in press).

\end{thebibliography}
\end{document}